\documentclass[aps,showpacs,preprintnumbers,amsmath,amssymb,superscriptaddress,floatfix,a4paper,nofootinbib,11pt,floatfix,nofootinbib,onecolumn]{revtex4}
\usepackage{graphicx}
\usepackage{bm}
\usepackage{rotating}
\usepackage{array}
\usepackage{xcolor}
\usepackage{amsmath,amssymb}
\usepackage{mathrsfs}
\usepackage{graphicx}
\usepackage{color}
\usepackage{subfigure}
\usepackage{fancyhdr}
\usepackage{multirow}
\usepackage{float}
\usepackage{epsfig}
\usepackage{amsfonts}
\usepackage{bm}
\usepackage{multirow}
\usepackage{nicefrac}
\newcommand{\ba}{\begin{eqnarray}}
\newcommand{\ea}{\end{eqnarray}}

\usepackage{bbm}
\newcommand{\be}{\begin{eqnarray}}
\newcommand{\ee}{\end{eqnarray}}
\usepackage{slashed}

\begin{document}

\title{Composite Inflation and further refining dS swampland conjecture}

\author{Jureeporn Yuennan} 
\email{jureeporn_yue@nstru.ac.th}
\affiliation{Surface Technology Research Unit (STRU), Faculty of Science and Technology, \\Nakhon Si Thammarat Rajabhat University, Nakhon Si Thammarat, 80280, Thailand}

\author{Phongpichit Channuie} 
\email{channuie@gmail.com}
\affiliation{College of Graduate Studies, Walailak University, Thasala, \\Nakhon Si Thammarat, 80160, Thailand}
\affiliation{School of Science, Walailak University, Thasala, \\Nakhon Si Thammarat, 80160, Thailand}

\date{\today}

\begin{abstract}

A natural combination of the first and second derivatives of the scalar potential was achieved in a framework of an alternative refined de Sitter conjecture recently proposed in the literature. In this work, we study various inflation models in which the inflaton is a composite field emerging from various strongly interacting field theories. We then examine if these three models of inflation can satisfy this further refining de Sitter swampland conjecture or not. Regarding our analysis with proper choices of parameters  $a,\,b=1- a$ and $q$, we find that some inflationary models are in strong tension with the refined Swampland conjecture. However, all of them can always satisfy the alternative refined de Sitter conjecture. Therefore, one may expect that all inflationary models might all be in “landscape” since the “further refining de Sitter swampland conjecture” is satisfied.

\end{abstract}


\maketitle


\section{Introduction}
The swampland program has been an exciting topic in cosmology and has been tested in various aspects such as black hole physics, inflation, dark energy, and late-time behavior. It has been originally developed for the phenomenology of quantum gravity theories.  A lot of work has been done recently about swampland conjectures, see a comprehensive review on the Swampland Refs.\cite{Brennan:2017rbf,Palti:2019pca}. Nevertheless, the original version have been adjusted to solve many cosmological problems. Recently, the refined version of the swampland conjecture has been suggested by the authors of Refs.\cite{Garg:2018reu,Ooguri:2018wrx}. In various phenomenological models, the topics include inflation \cite{Kinney:2018nny,Lin:2018rnx,Cheong:2018udx}, dark energy \cite{Agrawal:2018rcg,Chiang:2018lqx}, and other consequences \cite{Olguin-Trejo:2018zun,Junghans:2018gdb,Banlaki:2018ayh}. Yet, $H_{0}$ tension has also been discussed in the context of the Swampaland conjecture \cite{Banerjee:2020xcn,Colgain:2019joh}. Moreover it has been also discussed in relation to stringy constructions \cite{Garg:2018zdg,Blaback:2018hdo,Heckman:2018mxl,Blanco-Pillado:2018xyn,Emelin:2018igk} or in a more general swampland context \cite{Hebecker:2018vxz,Dvali:2018jhn,Schimmrigk:2018gch,Ibe:2018ffn,Yuennan:2022zml,Channuie:2019vsp,Gashti:2022pvu,NooriGashti:2022inp,NooriGashti:2022xmf}. In order to study inflation, we closely follow the work present in Ref.\cite{Andriot:2018mav}. We start by considering a four-dimensional ($4D$) theory of a (real) scalar field $\phi^{i}$ coupled to gravity. Hence its dynamics is governed by a scalar potential $V(\phi^{j})$. The
action takes the form
\begin{eqnarray}
S = \int_{4D} d^4x\sqrt{-g}\,\bigg[-\frac{1}{2}\,M_p^2\, R + \frac{1}{2}\,g^{\mu\nu}h_{ij}\partial_\mu \phi^{i}\,\partial_\nu \phi^{j} - V\bigg],\label{Ac}
\end{eqnarray}
where $h_{ij}(\phi^k)$ is the field space metric, $M_{p}$ is the 4D Planck mass, and the $4D$ space-time indexes $(\mu,\,\nu)$ are raised and lowered with the $4D$ metric $g_{\mu\nu}$ with a signature $(+,-,-,-)$. Various phenomenological models, such as multi-field cosmological inflation \cite{Dimopoulos:2005ac,Wands:2007bd}, can be described by the above action. As aforementioned, Swampland conjectures have been tested in various types of cosmological studies. The refined swampland conjecture, namely dS swampland, is govern by the following two conditions \cite{Garg:2018reu,Ooguri:2018wrx}:
\begin{eqnarray}
|\nabla V|\geq \frac{c_{1}}{M_{p}}V\,,\label{R1}
\end{eqnarray}
or
\begin{eqnarray}
{\rm min}(\nabla_{j}\nabla_{j} V)\,\,\leq -\frac{c_{2}}{M^{2}_{p}}V\,,\label{R2}
\end{eqnarray}
where $c_1$ and $c_2$ are both positive constants with $c_{1,2}\sim{\cal O}(1)$ and $|\nabla V|=\sqrt{g^{ij}\nabla_{j}\nabla_{j} V}$. Therefore,  for any $V$, the standard slow-roll parameters can be recast using the inequalities to yield
\begin{eqnarray}
\sqrt{2\varepsilon_{1}}\geq c_{1}\,,\quad{\rm or}\quad \varepsilon_{2}\,\,\leq -c_{2}\,.\label{R12}
\end{eqnarray}
The first condition corresponds to the original
“swampland conjecture” proposed in Ref.\cite{Obied:2018sgi}. However, a disconnected relation between these two distinct conditions (\ref{R1}) and (\ref{R2}) on two different quantities $\varepsilon_{1}$ and $\varepsilon_{2}$ was noticed. Therefore, a single condition on both $\varepsilon_{1}$ and $\varepsilon_{2}$ has been afterwards proposed dubbed a further
refining de Sitter swampland conjecture \cite{Andriot:2018mav}. 

The statement of this alternative refined de Sitter conjecture is suggested that at any point in field space a low energy effective theory of a quantum gravity that takes the form (\ref{Ac}) should satisfy where $V>0$ \cite{Andriot:2018mav}
\begin{eqnarray}
\Big(M_{p}\frac{|\nabla V|}{V}\Big)^{q}-aM^{2}_{p}\frac{{\rm min}(\nabla_{j}\nabla_{j} V)}{V}\geq b\,\quad{\rm with}\quad a+b=1,\,a,\,b>0,\,q>2\,,\label{ReRe}
\end{eqnarray}
which gives a natural combination of
the first and second derivatives of the scalar potential. Notice that the above inequality can be cast in terms of the slow-roll parameters. In terms of the slow-roll parameters, the conjecture can be rewritten as [1]:
\begin{eqnarray}
(2\varepsilon_{1})^{q/2}-a\varepsilon_{2}\geq b\,.\label{Re1}
\end{eqnarray}
We have found that the work of Ref.\cite{Liu:2021diz} has examined if Higgs inflation model, Palatini Higgs inflation, and Higgs-Dilaton model can satisfy the further refining de Sitter swampland conjecture or not. The author discovered that all inflationary models can always satisfy this new swampland conjecture if only they adjust the relevant parameters $a,\,b=1-a$ and $q$. 

In this work, various inflation models in which the inflaton is a composite field emerging from various strongly interacting field theories will be tested with the further refining de Sitter swampland conjecture. The paper is organized in the following way. In Section \ref{com}, we take a short recap on inflationary framework of which the inflaton is a composite field emerging from various strongly interacting field theories. For each model, the spectral index and the tensor-to-scalar ration will be derived. In Section \ref{s3}, we examine whether all
models satisfy the further refining swampland conjecture or not. We finally conclude our findings in the last section. 

\section{Composite Inflationary Scenario Revisited}\label{com}
In this section, we take a short recap on inflationary framework of which the inflaton is a composite field emerging from various strongly interacting field theories. 

\subsection{Composite NJL Inflation (NJLI)} 
The model has been proposed so far by the authors of Ref.\cite{Channuie:2016iyy}. In the inflation sector, the action describing model of inflation in which the inflaton is non-minimally coupled to gravity in the Jordan (J) frame takes the form
\begin{eqnarray} \label{NJLcurvedxi2}
S_{\rm{J}} &=& \int d^4x \sqrt{-g} \, \bigg[ - \frac{1}{2} M^{2}_{\textrm{\tiny{PL}}} R + \frac{1}{2}g^{\mu\nu}\partial_{\mu}\varphi\partial_{\nu}\varphi - \frac{\xi \, R}{2}\,  \left(\varphi^{2} - \frac{v^2}{2}\right)  - V_{J}(\varphi) \bigg]\,, \cr
V_{J}(\varphi) &=& - \frac{1}{2} m^{2}_{\varphi}\varphi^{2} + \frac{1}{2}\lambda\varphi^{4},
\end{eqnarray}
with $v$ being a vacuum expectation value (VEV) of the inflaton field $\varphi$. The above non-minimal coupling framework is actually inspired by Higgs-inflation investigated in Ref.\cite{Bezrukov:2007ep}. The importance is that the non-minimal coupling ($\xi$) of the Higgs doublet field $(H)$ to gravity, i.e. $\sim\xi H^{\dagger}HR$, is needed. With $\xi=0$, a realistic quartic Higgs self-interaction term generates an unacceptably large amplitude of primordial inhomogeneities \cite{Bezrukov:2008ut}. Specifically, with $\xi\sim 10^{4}$, the model leads to a successful inflation and produces a spectrum of primordial fluctuations in good agreement with the observational data \cite{Bezrukov:2007ep}.

Applying a conformal transformation, we can rewrite the action as minimally coupled but with a new canonically normalized field. Hence the conformal transformation can be basically implemented by making use of the following replacement:
\begin{equation}
\tilde{g}_{\mu\nu} = \Omega^{2}\,g_{\mu\nu} = \left(1 + \frac{\xi\,(\varphi^{2} - v^2/2)}{M^{2}_{p}}\right)g_{\mu\nu}\,.
\end{equation}
Therefore, the action in (\ref{NJLcurvedxi2}) becomes the Einstein-frame (E) form:
\begin{eqnarray} \label{NJLEif}
S_{\rm E} = \int d^4x \sqrt{-g} \, \bigg[ -\frac{1}{2} M^{2}_{p} R + \frac{1}{2}\Omega^{-4}\left(\Omega^{2} + \frac{6\xi\varphi^{2}}{M^2_{p}}\right)g^{\mu\nu}\partial_{\mu}\varphi\, \partial_{\nu}\varphi - U(\varphi)\bigg] \,,
\end{eqnarray}
where
\begin{equation} \label{UEi}
\Omega^{2} = \left(1 + \frac{\xi\,(\varphi^{2} - v^2/2)}{M^{2}_{p}}\right)\quad{\rm and}\quad U(\varphi)\equiv\Omega^{-4}V_{J}(\varphi)\,.
\end{equation}
Here, we end up with a non-canonical kinetic term for the scalar field. After introducing a new canonically normalized scalar field $\chi$, we have
\begin{eqnarray} \label{canonical}
\frac{1}{2}g^{\mu\nu}\partial_{\mu}\chi(\varphi)\partial_{\nu}\chi(\varphi) = \frac{1}{2}\left(\frac{d\chi}{d\varphi}\right)^{2}g^{\mu\nu}\partial_{\mu}\varphi\partial_{\nu}\varphi\,,
\end{eqnarray}
where
\begin{eqnarray} \label{fieldchi}
\chi^{\prime} = \left(\frac{d\chi}{d\varphi}\right) = \sqrt{\Omega^{-4}\left(\Omega^{2} + \frac{6\xi\varphi^{2}}{M^{2}_{p}}\right)}\,.
\end{eqnarray}
For small field value, i.e. $\xi\varphi^{2}\ll M^{2}_{p}$, the potential for the field $\chi$ becomes to that of the original field, $\varphi$. However, it is not the case for large-field value, i.e. $\xi\varphi^{2}\gg M^{2}_{p}$. In the later case, we can figure out the solution of $\varphi$ and  write in terms of the field $\chi$ as
\begin{eqnarray} \label{soluphi}
\varphi \simeq \frac{M_{p}}{\sqrt{\xi}}\exp\left(\frac{\chi}{\sqrt{6}M_{p}}\right)\,.
\end{eqnarray}
The effective potential $U(\chi)$ has the form
\begin{eqnarray} \label{solpot}
U(\chi) \simeq \frac{\lambda M^{4}_{p}}{2\xi^{2}}\bigg[1+ \exp\left(-\frac{2\chi}{\sqrt{6}M_{p}}\right)\bigg]^{-2}\,,
\end{eqnarray}
where we have assumed that the field is far away from the minimum of its potential such that $\xi v^{2}\ll M^{2}_{p}$. In the limit of $\varphi^{2}\gg M^{2}_{p}/\xi\gg v^{2}$, the slow-roll parameters in the Einstein frame can be written as functions of the field $\varphi(\chi)$:
\begin{eqnarray}
\varepsilon_{1}   & = & \frac{M^{2}_{p}}{2}\left(\frac{dU/d\chi}{U}\right)^{2} = \frac{M^{2}_{p}}{2}\left(\frac{U'}{U}\frac{1}{\chi'}\right)^{2} \simeq \frac{4M^{4}_{p}}{3\xi^{2}\varphi^{4}}, \cr
\varepsilon_{2} & = & M^{2}_{p}\frac{d^{2}U/d\chi^{2}}{U} = M^{2}_{\textrm{\tiny{PL}}}\frac{U''\chi' - \chi''}{U\chi'^{3}} \simeq -\frac{4M^{2}_{p}}{3\xi\varphi^{2}}\,,
\end{eqnarray}
where \lq\lq\,\,$'$\,\,\rq\rq\,denotes derivative with respect to $\varphi$. Notice that the results we obtained here are approximately the same for those of inflationary model driven by the SM Higgs boson \cite{Bezrukov:2007ep} and another composite inflation \cite{Channuie:2011rq}. We can quantify the field value at the end of inflation using $\epsilon = 1$. We find $\varphi_{\rm {\tiny{end}}} \simeq (4/3)^{1/4}M_{p}/\sqrt{\xi}$. The number of e-folds for the change of the field $\varphi$ from $\varphi_{N}$ to $\varphi_{\rm {\tiny{end}}}$ is given by
\begin{eqnarray} \label{efold}
N &=& \frac{1}{M^{2}_{p}} \int^{\chi_{N}}_{\chi_{\rm {\tiny{end}}}} \frac{U}{dU/d\chi}d\chi \nonumber\\&=& \frac{1}{M^{2}_{p}} \int^{\varphi_{N}}_{\varphi_{\rm {\tiny{end}}}} \frac{U}{dU/d\varphi}\left(\frac{d\chi}{d\varphi}\right)^{2}d\varphi \simeq \frac{6\xi}{8M^{2}_{p}}\left(\varphi^{2}_{N} - \varphi^{2}_{\rm {\tiny{end}}}\right)\,,
\end{eqnarray}
where $\varphi_{N}$ represents the field value at the horizon crossing of the observed CMB modes. After substituting $\varphi_{\rm {\tiny{end}}}$ into the above relation, we obtain $\varphi_{N} \simeq 9 M_{p}/\sqrt{\xi}$. To generate the proper amplitude of the density perturbations, the potential must satisfy the COBE renormalization $U/\epsilon \simeq (0.0276\,M_{p})^{4}$ \cite{Bezrukov:2008ut}. Inserting (\ref{UEi}) and (\ref{efold}) into the COBE normalization, we find the required value for $\xi$
\begin{eqnarray} \label{xi}
\xi \simeq \sqrt{\frac{2\lambda}{3}}\frac{N}{(0.0276)^{2}}\,.
\end{eqnarray} 

The spectral index of curvature perturbation $n_{s}$ and the tensor-to-scalar ratio $r$ are given in terms of the e-folds $N$:
\begin{eqnarray} \label{paras}
n_{s} & = & 1- 6\epsilon +2\eta \simeq 1-\frac{2}{N} - \frac{9}{2{N^{2}}}, \cr
r & = & 16\epsilon \simeq \frac{12}{N^{2}} .
\end{eqnarray}

For example, we obtain from Eq.(\ref{xi}) and (\ref{paras}) that $\xi\sim 64,000\sqrt{\lambda},\,n_{s}\simeq 0.965$ and $r\simeq 0.0033$ for $N=60$ e-folds.

\subsection{Glueball Inflation (GI)}
This class of model assumes that inflation is driven by a composite
state in a strongly interacting theory, see
\cite{Bezrukov:2011mv,Channuie:2012bv}. Let us consider the following model, see also Ref.\cite{Bezrukov:2011mv} for more details:
\begin{equation}
\label{eq:lagrangegiGI}
L_{GI}=\varphi^{-3/2}\partial_{\mu}\varphi\partial^{\mu}\varphi
-\frac{\varphi}{2}\ln\left(\frac{\varphi}{\Lambda^4}\right),
\end{equation}
where $\Lambda$ is a mass scale. Since the inflaton field in this case has mass dimension four, it is more convenient to work in the field $\phi$ with mass dimension one related to $\varphi$ via 
\begin{equation}
\frac{\varphi}{\Lambda^4}=\left(\frac{\phi}{\phi_{0}}\right)^4,
\end{equation}
with
\begin{equation}
\label{eq:defphizeroGI}
\phi_0=4\sqrt{2}\Lambda.
\end{equation}
Let us consider the situation where the model has a general non-minimal coupling to gravity of the form 
\begin{equation}
S =\int d^4x\sqrt{-g}\left[-\frac{M^2+\xi \Lambda^{2}\Big(\phi/\phi_{0}\Big)^{2}}{2}R+L_{GI}\right].
\end{equation}
The coupling to gravity is characterized by the parameter $\xi$. Then, the action in the Einstein frame
reads~\cite{Bezrukov:2011mv,Channuie:2012bv}
\begin{eqnarray}
S &=&\int d^4x \sqrt{-g}
\biggl[-\frac{1}{2}M^2_{p}R-
\Omega^{-2}\left(1+\frac{3\xi^2 \Lambda^{2}\Big(\phi/\phi_{0}\Big)^{2}}{16M_{p}^2}\Omega^{-2}
\right)\left(\frac{\Lambda}{\phi_{0}}\right)^{2}
\partial_{\mu}\phi\partial^{\mu}\phi
-\Omega^{-4}V_{\mathrm{GI}}\biggr],
\end{eqnarray}
where $V_{GI}$ refers to the potential in Eq.(\ref{eq:lagrangegiGI})
and $\Omega^2=\left(M^2+\xi\Lambda^{2}\big(\phi/\phi_{0}\big)^{2}\right)/M_{p}^2$. If $\xi\neq 0$ and if we are in the large field limit, then $\Omega^2\simeq\xi\Lambda^{2}\big(\phi/\phi_{0}\big)^{2}/M_{p}^2$. Therefore, we find
\begin{equation}
V_{GI} = 2\Lambda^4 \left( \frac{\phi}{\phi_{0}}\right)^4 \ln
\left(\frac{\phi}{\phi_{0}}\right),
\end{equation}
with $\phi_{0} \equiv 4\sqrt{2}\Lambda$. We now can introduce a canonically normalized field $\chi$ related to $\phi$ via
\begin{equation}
\frac{1}{2}{\tilde g}^{\mu\nu}\partial_{\mu}\chi(\phi)\partial_{\nu}\chi(\phi)=\frac{1}{2}\left(\frac{d\chi}{d\phi}\right)^{2}{\tilde g}^{\mu\nu}\partial_{\mu}\phi\partial_{\nu}\phi\,,
\end{equation}
where
\begin{equation}
\frac{1}{2}\left(\frac{d\chi}{d\phi}\right)^{2}=\Omega^{-2}\left(1+\frac{3\xi^2 \Lambda^{2}\Big(\phi/\phi_{0}\Big)^{2}}{16M_{p}^2}\Omega^{-2}
\right)\left(\frac{\Lambda}{\phi_{0}}\right)^{2}\,.
\end{equation}
The canonically normalized field $\chi$ is
such that $\chi  \propto \ln\phi$. In that case the potential
reduces to $\Omega^{-4}V_{GI}\propto \ln\phi$. Therefore, in terms of the canonically normalized field, we have:
\begin{eqnarray}
S_{E} &=&\int dx^4 \sqrt{-g}
\Bigg\{-\frac{1}{2}M_p^2 R+\frac{1}{2}g^{\mu\nu}\partial_{\mu}\chi\partial_{\nu}\chi-U_{GI}(\chi)\Bigg\}\,,
\end{eqnarray}
with
\begin{eqnarray}
U_{GI}(\chi)=\Omega^{-4}V_{GI}(\varphi)\,,\label{EUGI}
\end{eqnarray}
We now turn to the slow-roll analysis of the potential derived previously in Eq.(\ref{EUGI}). We consider here the large field regime $N_c^2\xi \Lambda^{2}\left(\frac{\phi}{\phi_{0}}\right)^{2}\gg M^{2}$. Hence the potential of this model
in the Einstein frame written in terms of the field $\phi$ takes the form
\begin{eqnarray}
U_{GI}(\phi)=\frac{2M_{p}^{4}}{\xi^{2}}\ln\Big(\frac{\phi}{\phi_{0}}\Big)\,.\label{EUphiGI}
\end{eqnarray}
with $\alpha$ being a constant which is expected to be of order unity \cite{Feo:2004mr}. It is rather straightforward to show that after inserting Eq.(\ref{EUphiGI}) into the slow-roll parameters, $\varepsilon_{i},\,i=1,2$, in this model, we find in the $\phi$ variable \cite{Channuie:2015ewa}
\begin{eqnarray}
\varepsilon_{1}&=&\frac{M^{2}_{p}}{2}\left(\frac{dU_{\rm GI}/d\chi}{U_{\rm GI}}\right)^{2}=\frac{M^{2}_{p}}{2}\left(\frac{dU_{\rm GI}/d\phi}{U_{GI}}\frac{d\phi}{d\chi}\right)^{2}\simeq \frac{4}{3\big(\ln(\phi/\phi_{0})\big)^{2}}\,,\\\varepsilon_{2}&=&M^{2}_{p}\left(\frac{d^{2}U_{GI}/d\chi^{2}}{U_{\rm GI}}\right)=M^{2}_{p}\left(\frac{dU_{\rm GI}/d\phi}{d\chi/d\phi}\right)'\left(\frac{d\phi/d\chi}{U_{\rm GI}}\right)=0
\,,\label{ep12GI}
\end{eqnarray}
where a prime denotes derivative with respect to the field $\phi$. In the above results, we have assumed the large non-minimal coupling $\xi\gg 1$. At the end of inflation, i.e., $\varepsilon_{1}(\phi_{end})=1$, we find for this model $\phi_{\rm end}/\phi_{0}=\exp(\sqrt{4/3})\sim 3.2$. In the large field limits, the number of e-foldings is given by
\begin{eqnarray}
N\simeq \frac{3}{16}\Big[\left(\ln\Big(\frac{\phi_{ini}}{\phi_{0}}\Big)\right)^{2}-\left(\ln\Big(\frac{\phi_{end}}{\phi_{0}}\Big)\right)^{2}\Big]\,.
\end{eqnarray}
The above relation yields
\begin{eqnarray}
\frac{\phi_{ini}}{\phi_{0}}\simeq \exp\left(\sqrt{16\,N/3}\right)\,.
\end{eqnarray}
Note that we have assume that $\phi_{ini}\gg \phi_{end}$ and the label {\it ini} signifies that the expression has to be evaluated at the beginning of the inflationary period. Further relevant information can be extracted using the WMAP \cite{Bezrukov:2008ut} normalization condition:
\begin{eqnarray}
\frac{U_{ini,\,{\rm SYM}}}{\varepsilon_{ ini}}=(0.0276\,M_{p})^{4}\,.
\end{eqnarray}
Using the above constraint, in the limit of $\xi\gg 1$, we can determine the magnitude of the non-minimal coupling which assumes the following value
\begin{eqnarray}
\xi\sim 4550\,N^{3/2}\,.
\end{eqnarray}
For this model, to the lowest order in the slow-roll approximation, the inflationary predictions in terms of the number
of e-foldings are given by
\begin{eqnarray}
n_{s}&=&1-6\varepsilon_{1}+2\varepsilon_{2}\simeq 1-\frac{3}{2\,N}\,.\\ r&=&16\varepsilon_{3}\simeq \frac{4}{N}\,.
\end{eqnarray}
Using $N=50\,(60)$, this model of composite inflation predicts $n_{s}\simeq 0.970\,(0.975)$ and $r\simeq 0.08\,(0.067)$.

\subsection{super Yang-Mills Inflation (SYMI)}
An instructive construction of the super Yang-Mills (SYM) effective Lagrangian was proposed in Ref.\cite{Veneziano:1982ah}, see also Ref.\cite{Sannino:2003xe}. We do not repeat it here. The model under consideration can be constructed using a $N=1$ supersymmetric Yang-Mills gauge theory. The underlying Lagrangian density can be written as 
\begin{eqnarray}
L=\frac{1}{4}F_{\mu \nu}^a F^{a\mu \nu}
+\frac{i}{2}\bar{\lambda}^a {\slashed D}_{ab}\lambda^b,
\end{eqnarray}
with $a=1, \cdots, N_c^2-1$, $N_c$ being the number characterizing
$SU(N_c)$ gauge group. $F_{\mu \nu}^a$ is the usual Yang-Mills field strength, $\lambda^a$ a spinor field and $\slashed{D}$ a covariant derivative. Note here that a composite scalar field in this case is a bound state denoted by $\varphi \simeq \lambda \bar{\lambda}$, which can actually emerge in the theory if a
strong interaction takes place. The effective Lagrangian aimed at describing its dynamics has been derived in \cite{Sannino:2003xe} and reads
\begin{equation}
L_{SYM} =-\frac{N_c^2}{\alpha}\left(\varphi\varphi^{\dagger}\right)^{-2/3}
\partial_{\mu}\varphi\partial^{\mu}\varphi^{\dagger}
-\frac{4\alpha N_c^2}{9}\left(\varphi\varphi^{\dagger}\right)^{2/3}
\ln\left(\frac{\varphi}{\Lambda^3}\right)
\ln\left(\frac{\varphi^{\dagger}}{\Lambda^3}\right),\label{pot}
\end{equation}
where $\alpha$ is a constant and $\Lambda$ a mass scale. Since the inflaton field in this case has mass dimension three, it is more convenient to work in the field $\phi$ with mass dimension one related to $\varphi$ via 
\begin{equation}
\frac{\varphi}{\Lambda^3}=\left(\frac{\phi}{\phi_{0}}\right)^3,
\end{equation}
with
\begin{equation}
\label{eq:defphizero}
\phi_0=3N_c\left(\frac{2}{\alpha}\right)^{1/2}\Lambda.
\end{equation}
and taking
$\varphi=\varphi^{\dagger}$. In the non-minimally coupled to gravity framework, e.g., Ref.\cite{Channuie:2012bv}, in the Jordan frame, we write
\begin{equation}
S_{J}=\int dx^4\sqrt{-g}\left[-\frac{M^2+N_c^2\xi
\Lambda^{2}\left(\frac{\phi}{\phi_{0}}\right)^{2}}{2}R+L_{SYM}\right],
\end{equation}
where $M$ is a mass scale. There is a new parameter in the problem, $\xi$, which describes the strength of the non-minimal coupling to gravity. Notice that a nonminimally-coupled inflaton sector to gravity was proposed in Higgs inflation \cite{Bezrukov:2007ep}. Then, in the Einstein frame, one can write the above model as \cite{Channuie:2012bv}
\begin{eqnarray}
S_{E} &=&\int dx^4 \sqrt{-g}
\Bigg\{-\frac{1}{2}M_p^2 R-\frac{9N_c^2}{\alpha}
\Omega^{-2}\left[1+\frac{\alpha N_c^2\xi^2}{3M_p^2}\Omega^{-2}
\Lambda^{2}\left(\frac{\phi}{\phi_{0}}\right)^{2}\right]
\left(\frac{\Lambda}{\phi_{0}}\right)^{2}
\partial_{\mu}\phi\partial^{\mu}\phi
\nonumber \\ & &
\quad\quad\quad\quad\quad-\Omega^{-4}V_{SYM}\Bigg\}.
\end{eqnarray}
In this expression, we refer $V_{SYM}$ as the second term in
Eq.~(\ref{pot}) and
\begin{equation}
\Omega^2\equiv \dfrac{M^2+N_c^2\xi
\Lambda^{2}\left(\frac{\phi}{\phi_{0}}\right)^{2}}{M^{2}_{p}}\,.
\end{equation}
In the following, we only consider a situation of which the case where $\xi\neq
0$ such that $\Omega ^2\simeq N_c^2\xi\Lambda^{2}\left(\frac{\phi}{\phi_{0}}\right)^{2}/M_p^2$, i.e., the second term in the definition of $\Omega^2$ dominates (the large field limit). In this case, we have
\begin{equation}
V_{SYM}(\phi)=4\alpha N^{2}_{c}\Lambda^{4}\left(\frac{\phi}{\phi_{0}}\right)^{4}\ln^{2} \left(\frac{\phi}{\phi_{0}}\right).
\end{equation}
We now can introduce a canonically normalized field $\chi$ related to $\phi$ via
\begin{equation}
\frac{1}{2}{\tilde g}^{\mu\nu}\partial_{\mu}\chi(\phi)\partial_{\nu}\chi(\phi)=\frac{1}{2}\left(\frac{d\chi}{d\phi}\right)^{2}{\tilde g}^{\mu\nu}\partial_{\mu}\phi\partial_{\nu}\phi\,,
\end{equation}
where
\begin{equation}
\frac{1}{2}\left(\frac{d\chi}{d\phi}\right)^{2}=\frac{9N_c^2}{\alpha}
\Omega^{-2}\left[1+\frac{\alpha N_c^2\xi^2}{3M_p^2}\Omega^{-2}
\Lambda^{2}\left(\frac{\phi}{\phi_{0}}\right)^{2}\right]
\left(\frac{\Lambda}{\phi_{0}}\right)^{2}\,.
\end{equation}
Therefore, in terms of the canonically normalized field, we have:
\begin{eqnarray}
S_{E} &=&\int dx^4 \sqrt{-g}
\Bigg\{-\frac{1}{2}M_p^2 R+\frac{1}{2}g^{\mu\nu}\partial_{\mu}\chi\partial_{\nu}\chi-U_{SYM}(\chi)\Bigg\}\,,
\end{eqnarray}
with
\begin{eqnarray}
U_{SYM}(\chi)=\Omega^{-4}V_{SYM}(\phi)\,,\label{EU}
\end{eqnarray}
We now turn to the slow-roll analysis of the potential derived previously in Eq.(\ref{EU}). We consider here the large field regime $N_c^2\xi \Lambda^{2}\left(\frac{\phi}{\phi_{0}}\right)^{2}\gg M^{2}$. Hence the potential of this model
in the Einstein frame written in terms of the field $\phi$ takes the form
\begin{eqnarray}
U_{SYM}(\phi)=\frac{4\alpha}{N^{2}_{c}}\frac{M_{p}^{4}}{\xi^{2}}\ln^{2}\left(\frac{\phi}{\phi_{0}}\right)\,.\label{EUphi}
\end{eqnarray}
with $\alpha$ being a constant
which is expected to be of order unity \cite{Feo:2004mr}. It is rather straightforward to show that after inserting Eq.(\ref{EUphi}) into the slow-roll parameters, $\varepsilon_{i},\,i=1,2$, in this model, we find in the $\phi$ variable \cite{Channuie:2015ewa}
\begin{eqnarray}
\varepsilon_{1}&=&\frac{M^{2}_{p}}{2}\left(\frac{dU_{\rm SYM}/d\chi}{U_{\rm SYM}}\right)^{2}=\frac{M^{2}_{p}}{2}\left(\frac{dU_{\rm SYM}/d\phi}{U_{\rm SYM}}\frac{d\phi}{d\chi}\right)^{2}\simeq \frac{1}{3\big(\ln(\phi/\phi_{0})\big)^{2}}\,,\\\varepsilon_{2}&=&M^{2}_{p}\left(\frac{d^{2}U_{\rm SYM}/d\chi^{2}}{U_{\rm SYM}}\right)=M^{2}_{p}\left(\frac{dU_{\rm SYM}/d\phi}{d\chi/d\phi}\right)'\left(\frac{d\phi/d\chi}{U_{\rm SYM}}\right)\simeq \frac{1}{3\big(\ln(\phi/\phi_{0})\big)^{2}}
\,,\label{ep12}
\end{eqnarray}
where a prime denotes derivative with respect to the field $\phi$. In the above results, we have assumed the large non-minimal coupling $\xi\gg 1$. At the end of inflation, i.e., $\varepsilon_{1}(\phi_{end})=1$, we find for this model $\phi_{\rm end}/\phi_{0}=\exp(\sqrt{1/3})\sim 1.8$. In the large field limits, the number of e-foldings is given by
\begin{eqnarray}
N\simeq \frac{3}{2}\Big[\left(\ln\Big(\frac{\phi_{ini}}{\phi_{0}}\Big)\right)^{2}-\left(\ln\Big(\frac{\phi_{end}}{\phi_{0}}\Big)\right)^{2}\Big]\,.
\end{eqnarray}
The above relation yields
\begin{eqnarray}
\frac{\phi_{ini}}{\phi_{0}}\simeq \exp\left(\sqrt{2N/3}\right)\,.
\end{eqnarray}
Note that we have assume that $\phi_{ini}\gg \phi_{end}$ and the label {\it ini} signifies that the expression has to be evaluated at the beginning of the inflationary period. Further relevant information can be extracted using the WMAP \cite{Bezrukov:2008ut} normalization condition:
\begin{eqnarray}
\frac{U_{ini,\,{\rm SYM}}}{\phi_{ ini}}=(0.0276\,M_{p})^{4}\,.
\end{eqnarray}
Using the above constraint, in the limit of $\xi\gg 1$ we find
\begin{eqnarray}
N_{c}\xi\sim 3000\sqrt{\alpha}\,N\,.
\end{eqnarray}
We see that it is possible to lower the value of a nonminimal coupling $\xi$ by increasing the number of underlying colors, $N_{c}$. We recall that $\alpha$ is given by the underlying theory and is expected to be of order unity \cite{Feo:2004mr}. For this model, to the lowest order in the slow-roll approximation, the inflationary predictions in terms of the number
of e-foldings are given by
\begin{eqnarray}
n_{s}&=&1-6\varepsilon_{1}+2\varepsilon_{2}\simeq 1-\frac{2}{N}\,.\\ r&=&16\varepsilon_{1}\simeq \frac{8}{N}\,.
\end{eqnarray}
Using $N=60$, this model of composite inflation predicts $n_{s}\simeq 0.967$ and $r\simeq 0.133$. However, it is intention with a tightened value obtained from a combination of the BICEP2/Keck Array BK15 data of $r_{0.002}<0.056$. 

\subsection{Orientifold Inflation (OI)}
This class of theories are discussed in more detail in xxxxx. However, in Ref.\cite{Channuie:2012bv}, it was argued that in ``orientifold theories'', the above Lagrangian can be slightly deformed and now takes the form
\begin{equation}
\label{eq:lagrangeoiOI}
L_{OI} =-\frac{N_c^2}{\alpha_{OI}}\left(\varphi\varphi^{\dagger}\right)^{-2/3}
\partial_{\mu}\varphi\partial^{\mu}\varphi^{\dagger}
-\frac{4\alpha_{OI} N_c^2}{9}\left(\varphi\varphi^{\dagger}\right)^{2/3}
\left[\ln\left(\frac{\varphi}{\Lambda^3}\right)
\ln\left(\frac{\varphi^{\dagger}}{\Lambda^3}\right)
-\beta\right],
\end{equation}
where $\beta={\cal O}(1/N_c)$. Similarly, we can follow our analysis given in the previous subsection. We take the field $\phi$ with mass dimension one related to $\varphi$ via 
\begin{equation}
\frac{\varphi}{\Lambda^3}=\left(\frac{\phi}{\phi_{0}}\right)^3,
\end{equation}
with
\begin{equation}
\label{eq:defphizeroOI}
\phi_0=3N_c\left(\frac{2}{\alpha}\right)^{1/2}\Lambda.
\end{equation}
and taking
$\varphi=\varphi^{\dagger}$. Having considered the non-minimally coupled to gravity framework, we write in the Jordan frame
\begin{equation}
S_{J}=\int dx^4\sqrt{-g}\left[-\frac{M^2+N_c^2\xi
\Lambda^{2}\left(\frac{\phi}{\phi_{0}}\right)^{2}}{2}R+L_{OI}\right],
\end{equation}
where $M$ is a mass scale. Again the new parameter, $\xi$, describes the strength of the non-minimal coupling to gravity. Then, in the Einstein frame, one can write the above model as \cite{Channuie:2012bv}
\begin{eqnarray}
S_{E} &=&\int dx^4 \sqrt{-g}
\Bigg\{-\frac{1}{2}M_p^2 R-\frac{9N_c^2}{\alpha}
\Omega^{-2}\left[1+\frac{\alpha N_c^2\xi^2}{3M_p^2}\Omega^{-2}
\Lambda^{2}\left(\frac{\phi}{\phi_{0}}\right)^{2}\right]
\left(\frac{\Lambda}{\phi_{0}}\right)^{2}
\partial_{\mu}\phi\partial^{\mu}\phi
\nonumber \\ & &
\quad\quad\quad\quad\quad-\Omega^{-4}V_{OI}\Bigg\}.
\end{eqnarray}
In this expression, we refer $V_{OI}$ as the second term in
Eq.~(\ref{eq:lagrangeoiOI}) and
\begin{equation}
\Omega^2\equiv \dfrac{M^2+N_c^2\xi
\Lambda^{2}\left(\frac{\phi}{\phi_{0}}\right)^{2}}{M^{2}_{p}}\,.
\end{equation}
In the following, we only consider a situation where $\xi\neq
0$ and then the second term in the definition of $\Omega^2$ dominates (the large field limit). In this case, we find
\begin{equation}
V_{OI}(\phi)=4\alpha N^{2}_{c}\Lambda^{4}\left(\frac{\phi}{\phi_{0}}\right)^{4}\Big[\ln^{2} \left(\frac{\phi}{\phi_{0}}\right)-\frac{\beta}{9}\Big].
\end{equation}
We now can introduce a canonically normalized field $\chi$ related to $\varphi$ via
\begin{equation}
\frac{1}{2}{\tilde g}^{\mu\nu}\partial_{\mu}\chi(\phi)\partial_{\nu}\chi(\phi)=\frac{1}{2}\left(\frac{d\chi}{d\phi}\right)^{2}{\tilde g}^{\mu\nu}\partial_{\mu}\phi\partial_{\nu}\phi\,,
\end{equation}
where
\begin{equation}
\frac{1}{2}\left(\frac{d\chi}{d\phi}\right)^{2}=\frac{9N_c^2}{\alpha}
\Omega^{-2}\left[1+\frac{\alpha N_c^2\xi^2}{3M_p^2}\Omega^{-2}
\Lambda^{2}\left(\frac{\phi}{\phi_{0}}\right)^{2}\right]
\left(\frac{\Lambda}{\phi_{0}}\right)^{2}\,.
\end{equation}
Therefore, in terms of the canonically normalized field, we have:
\begin{eqnarray}
S_{E} &=&\int dx^4 \sqrt{-g}
\Bigg\{-\frac{1}{2}M_p^2 R+\frac{1}{2}g^{\mu\nu}\partial_{\mu}\chi\partial_{\nu}\chi-U_{OI}(\chi)\Bigg\}\,,
\end{eqnarray}
with
\begin{eqnarray}
U_{OI}(\chi)=\Omega^{-4}V_{OI}(\varphi)\,,\label{EUOI}
\end{eqnarray}
We now use the slow-roll analysis of the potential derived previously in Eq.(\ref{EU}). We consider here the large field regime $N_c^2\xi \Lambda^{2}\left(\frac{\phi}{\phi_{0}}\right)^{2}\gg M^{2}$. Hence the potential of this model
in the Einstein frame written in terms of the field $\phi$ takes the form
\begin{eqnarray}
U_{OI}(\phi)=\frac{4\alpha}{N^{2}_{c}}\frac{M_{p}^{4}}{\xi^{2}}\Big[\ln^{2}\left(\frac{\phi}{\phi_{0}}\right)-\frac{\beta}{9}\Big]\,.\label{EUphiOI}
\end{eqnarray}
Notice that at large $N_c$ limit, i.e., $\beta\rightarrow 0$,
this theory maps into the preceding one. It is rather straightforward to show that after inserting Eq.(\ref{EUphiOI}) into the slow-roll parameters, $\varepsilon_{i},\,i=1,2$, in this model, we find for the large field approximation \cite{Channuie:2015ewa}
\begin{eqnarray}
\varepsilon_{1}=\varepsilon_{2}&=&\frac{M^{2}_{p}}{2}\left(\frac{dU_{\rm OI}/d\chi}{U_{\rm OI}}\right)^{2}=\frac{M^{2}_{p}}{2}\left(\frac{dU_{\rm OI}/d\phi}{U_{OI}}\frac{d\phi}{d\chi}\right)^{2}\nonumber\\&\simeq& \frac{1}{3\big(\ln(\phi/\phi_{0})\big)^{2}}\Big(1+\frac{2\beta}{9\big(\ln(\phi/\phi_{0})\big)^{2}}\Big)\,,\label{ep12OI}
\end{eqnarray}
where a prime denotes derivative with respect to the field $\phi$. In the above results, we have assumed the large non-minimal coupling $\xi\gg 1$. At the end of inflation, i.e., $\varepsilon_{1}(\phi_{end})=1$, we find for this model $\phi_{\rm end}/\phi_{0}=\exp(\sqrt{1/3})(1+\beta/(3\sqrt{3}))\sim 1.8(1+0.2\beta)$. In the large field limits, the number of e-foldings is given by
\begin{eqnarray}
N\simeq \frac{3}{2}\Bigg[\left(\ln\Big(\frac{\phi}{\phi_{0}}\Big)\right)^{2}\Big(1-\frac{2\ln\ln(\phi/\phi_{0})}{81\left(\ln\big(\phi/\phi_{0}\big)\right)^{2}}\beta\Big)\Bigg]\Bigg|^{\phi=\phi_{ini}}_{\phi=\phi_{end}}\,.
\end{eqnarray}
The above relation yields
\begin{eqnarray}
\frac{\phi_{ini}}{\phi_{0}}\simeq \exp\left(\sqrt{2N/3}\right)\Bigg(1+\Big[4+3\ln(2N)\Big]\frac{\beta}{12\sqrt{6N}}\Bigg)\,.
\end{eqnarray}
Note that we have assume that $\phi_{ini}\gg \phi_{end}$ and the label {\it ini} signifies that the expression has to be evaluated at the beginning of the inflationary period. For this model, to the lowest order in the slow-roll approximation, the inflationary predictions in terms of the number
of e-foldings are given by
\begin{eqnarray}
n_{s}&=&1-6\varepsilon_{1}+2\varepsilon_{2}\simeq 1-\frac{2}{N}-\frac{2\beta}{3N^{2}}\,.\\ r&=&16\varepsilon_{1}\simeq \frac{8}{N}+\frac{8\beta}{3N^{2}}\,,
\end{eqnarray}
with $\beta$ being a numerical (real) parameter with $\beta\sim {\cal O}(1/N_{c})$.

\section{Examination with the further refining swampland conjecture}\label{s3}
In this section, we closely follow Ref.\cite{Andriot:2018mav} of which a natural condition on a combination of the first and second derivatives of the scalar potential is achieved. Let us first define two new parameters for any scalar field $V(\phi)$ to yield
\begin{eqnarray}
F_{1}=\frac{|dV(\phi)/d\phi|}{V(\phi)}\,,
\label{f1}
\end{eqnarray}
and
\begin{eqnarray}
F_{2}=\frac{d^{2}V(\phi)/d\phi^{2}}{V(\phi)}\,.
\label{f2}
\end{eqnarray}
Considering Eq.(\ref{R12}), the above two parameters can be then rewritten in terms of the slow-roll parameters. Here we have
\begin{eqnarray}
F_{1}=\sqrt{2\varepsilon_{1}}\,,\quad F_{2}=\varepsilon_{2}\,.
\label{f1f2}
\end{eqnarray}
It is very useful to express $F_1$ and $F_2$ in terms of the slow-roll parameters. Therefore, we can relate them to the spectrum index of the primordial curvature power spectrum $n_{s}$ and tensor-to-scalar ratio $r$. In this case, it is rather straightforward to show that
\begin{eqnarray}
F_{1}=\sqrt{2\varepsilon_{1}}=\sqrt{\frac{r}{8}}\,,
\label{f11}
\end{eqnarray}
and
\begin{eqnarray}
F_{2}=\varepsilon_{2}=\frac{1}{2}\big(n_{s}-1+3r/8\big)\,.
\label{f12}
\end{eqnarray}
We will consider three models of inflation in which the inflaton is a composite field emerging from different underlying strongly interacting theories. We test the models if they satisfy this new refined swampland conjecture, or not.

\subsection{Composite NJL Inflation (NJLI)}
Using $n_{s}=0.965$, we solve for this model to obtain $N\sim 60$ implying that $r\sim 0.032<0.10$. We see that this value of $r$ is good agreement with a tightened value obtained from a combination of the BICEP2/Keck Array BK15 data of $r_{0.002}<0.056$. Inserting these values into Eq.(\ref{f1}) and Eq.(\ref{f2}), we obtain
\begin{eqnarray}
F_{1}&=&\sqrt{2\varepsilon_{1}}=\sqrt{\frac{r}{8}}=0.0204124\,,\\F_{2}&=&\varepsilon_{2}=\frac{1}{2}\big(n_{s}-1+3r/8\big)=-0.016875\,.
\label{f1f2}
\end{eqnarray}
Considering the refined swampland conjecture (\ref{R12}), we find
\begin{eqnarray}
c_{1}\leq 0.0204124\quad{\rm or}\quad c_{2}\leq 0.016875\,.
\label{c1c21n}
\end{eqnarray}
Clearly, $c_{1}$ and $c_{2}$ are both not ${\cal O}(1)$, implying that this composite model is in strong tension with the refined swampland conjecture. Let us next examine if the model satisfies the refining swampland conjecture. Using Eq.(\ref{ReRe}), we find
\begin{eqnarray}
(2\varepsilon_{1})^{q/2}-a\,\varepsilon_{2}\geq 1-a\,,\quad q>2\,.
\label{re1110n}
\end{eqnarray}
Substituting Eq.(\ref{c1c21n}) into Eq.(\ref{re1110n}), we find
\begin{eqnarray}
0.0204124^{q}+0.016875\,a\geq 1-a\,\quad{\rm or}\quad 0.107833^{q}\geq 1-1.016880\,a\,.
\label{re111}
\end{eqnarray}
If we can find $a$ to satisfy the condition
\begin{eqnarray}
\frac{1}{1.016880}(1-0.0204124^{q})\leq a<1\,,
\quad q>2,
\label{r111}
\end{eqnarray}
then the further refining swampland conjecture can be satisfied. In this case, when $a=\nicefrac{1}{1.016880}$, we have $1-1.016880\,a=0$. Therefore, we can examine that when $a < \nicefrac{1}{1.016880}$, we can always find a $q$ whose value is larger than $2$. It is possible to give an example of values of the parameters $a,\,b,\,q$, which work for this model. From Eq.(\ref{r111}), we use $q=3.0$ which is satisfied by a condition $q>2$. We find for this particular case that $0.970000 \leq a<1$ and choose $a=0.970000<\nicefrac{1}{1.016880}=0.983405$ and $1-a=1-0.97000=b=0.03000>0$.

\subsection{Glueball Inflation (GI)}
Using $n_{s}=0.965$, we can solve for this model to obtain $N\sim 43$ implying that $r\sim 0.09<0.10$. We see that this value of $r$ is intention with a tightened value obtained from a combination of the BICEP2/Keck Array BK15 data of $r_{0.002}<0.056$. Inserting these values into Eq.(\ref{f1}) and Eq.(\ref{f2}), we obtain
\begin{eqnarray}
F_{1}&=&\sqrt{2\varepsilon_{V}}=\sqrt{\frac{r}{8}}=0.107833\,,\\F_{2}&=&\eta_{V}=\frac{1}{2}\big(n_{s}-1+3r/8\big)=-0.0000581395\,.
\label{f1f2}
\end{eqnarray}
Considering the refined swampland conjecture (\ref{R12}), we find
\begin{eqnarray}
c_{1}\leq 0.107833\quad{\rm or}\quad c_{2}\leq 0.0000581395\,.
\label{c1c21}
\end{eqnarray}
Clearly, $c_{1}$ and $c_{2}$ are both not ${\cal O}(1)$, implying that this composite model is in strong tension with the refined swampland conjecture. Let us next examine if the model satisfies the refining swampland conjecture. Taking Eq.(\ref{ReRe}), we find
\begin{eqnarray}
(2\epsilon_{V})^{q/2}-a\eta_{V}\geq 1-a\,,\quad q>2\,.
\label{re1110}
\end{eqnarray}
Substituting Eq.(\ref{c1c21}) into Eq.(\ref{re1110}), we find
\begin{eqnarray}
0.107833^{q} +0.0000581395a\geq 1-a\,\quad{\rm or}\quad 0.107833^{q}\geq 1-1.0000581395\,a\,.
\label{re111}
\end{eqnarray}
If we can find $a$ to satisfy the condition
\begin{eqnarray}
\frac{1}{1.0000581395}(1-0.107833^{q})\leq a<1\,,
\quad q>2,
\label{r111}
\end{eqnarray}
then the further refining swampland conjecture can be satisfied. In this case, when $a=\nicefrac{1}{1.0000581395}$, we have $1-1.0000581395\,a=0$. Therefore, we can examine that when $a < \nicefrac{1}{1.0000581395}$, we can always find a $q$ whose value is larger than $2$. It is possible to give an example of values of the parameters $a,\,b,\,q$, which work for this model. From Eq.(\ref{r111}), we use $q=2.5$ which is satisfied by a condition $q>2$. We find for this particular case that $0.996124 \leq a<1$ and choose $a=0.980000<\nicefrac{1}{1.0000581395}=0.996124$ and $1-a=1-0.98000=b=0.02000>0$.

\subsection{super Yang-Mills Inflation (SYMI)}
Using $n_{s}=0.965$, we can solve for this model to obtain $N\sim 57$ implying that $r\sim 0.14 \nless 0.10$. We see that this value of $r$ is also intention with a tightened value obtained from a combination of the BICEP2/Keck Array BK15 data of $r_{0.002}<0.056$. In order to be satisfied to this upper bound, this model requires a large number of e-folds. Taking $r<0.056$, we find that $N>140$ for this model. Inserting $r\sim 0.14$ into Eq.(\ref{f1}) and Eq.(\ref{f2}), we obtain
\begin{eqnarray}
F_{1}&=&\sqrt{2\varepsilon_{V}}=\sqrt{\frac{r}{8}}=0.132453\,,\\F_{2}&=&\eta_{V}=\frac{1}{2}\big(n_{s}-1+3r/8\big)=0.00881579\,.
\label{f1f2}
\end{eqnarray}
Considering the refined swampland conjecture (\ref{R12}), we find
\begin{eqnarray}
c_{1}\leq 0.132453\quad{\rm or}\quad c_{2}\leq -0.00881579\,.
\label{c1c212}
\end{eqnarray}
Clearly, $c_{1}$ and $c_{2}$ are both not ${\cal O}(1)$, implying that this composite model is in strong tension with the refined swampland conjecture. We also noticed that $c_{2}$ is not positive. Let us next examine if the model satisfies the refining swampland conjecture. Taking Eq.(\ref{ReRe}), we find
\begin{eqnarray}
(2\epsilon_{V})^{q/2}-a\eta_{V}\geq 1-a\,,\quad q>2\,.
\label{re11102}
\end{eqnarray}
Substituting Eq.(\ref{c1c212}) into Eq.(\ref{re11102}), we find
\begin{eqnarray}
0.132453^{q} -0.00881579a\geq 1-a\,\quad{\rm or}\quad 0.107833^{q}\geq 1-0.99118\,a\,.
\label{re111}
\end{eqnarray}
If we can find $a$ to satisfy the condition
\begin{eqnarray}
\frac{1}{0.99118}(1-0.132453^{q})\leq a<1\,,
\quad q>2,
\label{r111}
\end{eqnarray}
then the further refining swampland conjecture can be satisfied. In this case, when $a=\nicefrac{1}{0.99118}$, we have $1-0.99118\,a=0$. Therefore, we can examine that when $a < \nicefrac{1}{0.99118}$, we can always find a $q$ whose value is larger than $2$. It is possible to give an example of values of the parameters $a,\,b,\,q$, which work for this model. From Eq.(\ref{r111}), we use $q=2.2$ which is satisfied by a condition $q>2$. We find for this particular case that $0.997085 \leq a<1$ and choose $a=0.980000<\nicefrac{1}{0.99118}=1.0089$ and $1-a=1-0.98000=b=0.02000>0$.

\subsection{Orientifold Inflation (OI)}
In our analysis below, we will take $N_{c}=3$ and then find $\beta\sim 1/3\sim 0.3$. Using $n_{s}=0.965$, we solve for this model to obtain $N\sim 57$ and then find that $r\sim 0.140\nless 0.10$. Therefore, this value of $r$ is in strong tension with a tightened value obtained from a combination of the BICEP2/Keck Array BK15 data of $r_{0.002}<0.056$. Inserting these values into Eq.(\ref{f1}) and Eq.(\ref{f2}), we obtain
\begin{eqnarray}
F_{1}&=&\sqrt{2\varepsilon_{1}}=\sqrt{\frac{r}{8}}=0.132569\,,\\F_{2}&=&\varepsilon_{2}=\frac{1}{2}\big(n_{s}-1+3r/8\big)=0.00735669\,.
\label{f1f2}
\end{eqnarray}
Considering the refined swampland conjecture (\ref{R12}), we find
\begin{eqnarray}
c_{1}\leq 0.132569\quad{\rm or}\quad c_{2}\leq -0.00735669\,.
\label{c1c21o}
\end{eqnarray}
Clearly, $c_{1}$ and $c_{2}$ are both not ${\cal O}(1)$, implying that this composite model is in strong tension with the refined swampland conjecture. Let us next examine if the model satisfies the refining swampland conjecture. Using Eq.(\ref{ReRe}), we find
\begin{eqnarray}
(2\varepsilon_{1})^{q/2}-a\,\varepsilon_{2}\geq 1-a\,,\quad q>2\,.
\label{re1110o}
\end{eqnarray}
Substituting Eq.(\ref{c1c21o}) into Eq.(\ref{re1110o}), we find
\begin{eqnarray}
0.132569^{q} - 0.00735669\,a\geq 1-a\,\quad{\rm or}\quad 0.107833^{q}\geq 1-0.992643\,a\,.
\label{re111}
\end{eqnarray}
If we can find $a$ to satisfy the condition
\begin{eqnarray}
\frac{1}{0.992643}(1-0.132569^{q})\leq a<1\,,
\quad q>2,
\label{r111}
\end{eqnarray}
then the further refining swampland conjecture can be satisfied. In this case, when $a=\nicefrac{1}{0.992643}$, we have $1-0.992643\,a=0$. Therefore, we can examine that when $a < \nicefrac{1}{0.992643}$, we can always find a $q$ whose value is larger than $2$. It is possible to give an example of values of the parameters $a,\,b,\,q$, which work for this model. From Eq.(\ref{r111}), we use $q=2.1$ which is satisfied by a condition $q>2$. We find for this particular case that $0.98000 \leq a<1$ and choose $a=0.99000<\nicefrac{1}{0.992643}=1.00741$ and $1-a=1-0.99000=b=0.01000>0$.

\section{Conclusion}
In general, the space of effective theories can be consistently coupled to a theory of quantum gravity. However, it can include the set of phenomenological models which cannot be derived as a low energy effective theory of a quantum gravity in the high energy regime. Therefore, it is required for consistent EFTs not to lie in the swampland. Swampland conjecture has gained significant attention which allows us to validate or invalidate a large class of low energy effective theories. However, a natural combination of the first and second derivatives of the scalar potential was recently achieved in a framework of an alternative refined de Sitter conjecture recently proposed in the literature. 

In this work, we studied various inflation models in which the inflaton is a composite field emerging from various strongly interacting field theories. We examined if these three models of inflation can satisfy this further refining de Sitter swampland conjecture or not. We discussed the theoretical viability of various composite models of inflation in light of the recent refined Swampland conjectures. Regarding our analysis with proper choices of parameters  $a,\,b=1- a$ and $q$, we discovered that some inflationary models are in strong tension with the refined Swampland conjecture. However, all of them always satisfied with the alternative refining de Sitter conjecture. Therefore, it was expected that all inflationary models might all be in “landscape” since the “further refining de Sitter swampland conjecture” is satisfied.

\acknowledgments
P. Channuie acknowledged the Mid-Career Research Grant 2020 from National Research Council of Thailand (NRCT5-RSA63019-03).

\end{document}